\pdfoutput=1
\documentclass[a4paper,USenglish]{lipics-v2021}

\hideLIPIcs  

\newcommand{\calS}{\mathcal{S}}

\newcommand{\calG}{\mathcal{G}}
\newcommand{\calF}{\mathcal{F}}

\newcommand{\temph}[1]{\textbf{#1}}

\usepackage[most]{tcolorbox}

\newcommand{\mypara}[1]{\smallskip\noindent\textbf{#1.}}

\title{Grassroots Platforms with Atomic Transactions: 
Social Networks, Cryptocurrencies, and Democratic Federations}

\titlerunning{Grassroots Platforms with Atomic Transactions} 

\author{Ehud {Shapiro}}{Weizmann Institute of Science, Israel \and London School of Economics, UK}{}{}{} 

\authorrunning{Ehud Shapiro} 

\Copyright{Ehud Shapiro} 



\relatedversion{} 

\nolinenumbers 

\EventEditors{John Q. Open and Joan R. Access}
\EventNoEds{2}
\EventLongTitle{42nd Conference on Very Important Topics (CVIT 2016)}
\EventShortTitle{CVIT 2016}
\EventAcronym{CVIT}
\EventYear{2016}
\EventDate{December 24--27, 2016}
\EventLocation{Little Whinging, United Kingdom}
\EventLogo{}
\SeriesVolume{42}

\begin{document}

\maketitle

\begin{abstract}
Grassroots platforms aim to offer an egalitarian alternative to global platforms --- centralized/autocratic (Facebook etc.) and decentralized/plutocratic (Bitcoin etc.) alike.
Whereas global platforms can have only a single instance---one Facebook, one Bitcoin---grassroots platforms can have multiple instances that emerge and operate independently of each other and of any global resource except the network, and can interoperate and coalesce into ever-larger instances once interconnected, potentially (but not necessarily) forming a single instance.
Key grassroots platforms include grassroots social networks,  grassroots cryptocurrencies, and grassroots democratic federations. Previously,  grassroots platforms were defined formally and proven grassroots using unary distributed transition systems, in which each transition is carried out by a single agent. 
However, grassroots platforms cater for a more abstract specification using transactions carried out atomically by multiple agents, something that cannot be expressed by unary transition systems.  
As a result, their original specifications and proofs were unnecessarily cumbersome and opaque.

Here, we aim to provide a more suitable formal foundation for grassroots platforms.  To do so, we enhance the notion of a distributed transition system to include atomic transactions and revisit the notion of grassroots platforms within this new foundation.    We present crisp specifications of key grassroots platforms using atomic transactions: befriending and unfriending for grassroots social networks, coin swaps for grassroots cryptocurrencies, and communities forming, joining, and leaving a federation for grassroots democratic federations.  We prove a general theorem that a platform specified by atomic transactions that are so-called interactive is grassroots; show that the atomic transactions used to specify all three platforms  are interactive; and conclude that the platforms thus specified are indeed grassroots.  We thus provide a crisp mathematical foundation for grassroots platforms and a solid and clear starting point from which their implementation can commence.
\end{abstract}

\section{Introduction}

\mypara{Background} The Internet today is dominated by centralised global platforms---social networks, Internet commerce, ‘sharing-economy’---with autocratic control~\cite{zuboff2019age,zuboff2022surveillance}.  An alternative is emerging---blockchains and cryptocurrencies~\cite{ethereum:dao,faqir2021comparative,nakamoto2008peer,ethereum,wang2018overview,wang2021ethereum}---that are also global platforms, but with decentralized, plutocratic control~\cite{vitalikplutocracy}.

Grassroots platforms~\cite{shapiro2023grassrootsBA,shapiro2023gsn,shapiro2024gc,halpern2024federated} aim offer an egalitarian alternative to global platforms, centralized and decentralized alike.  Global platforms can only have a single instance--one Facebook, one Bitcoin---as two instances of the platform would clash over the global resources they utilize---domain name, port number, or boot nodes, which are hardwired into their code.  Whether it is possible to modify their code (`hard-fork') to create non-conflicting instances---Facebook$'$, Bitcoin$'$---such forked instances would ignore, rather than interoperate with, the primary instance. 
Grassroots platforms, in contrast, can have multiple instances that emerge and operate independently of each other and of any global resource except the network, yet can interoperate and coalesce once interconnected, forming ever-larger platform instances, potentially (but not necessarily) coalescing into a single instance. 

Any platform that operates on a shared global resource or employs a single replicated (Blockchain~\cite{bitcoin}), or distributed (IPFS~\cite{benet2014ipfs}, DHT~\cite{rhea2005opendht}) shared global data structure, and distributed pub/sub systems with a global directory~\cite{chockler2007constructing,chockler2007spidercast,buchegger2009peerson}, are all not grassroots.    
BitTorrent~\cite{bittorrent} and Mastodon~\cite{raman2019challenges} are similar in spirit but are peer-to-peer among servers not people; in particular, in such systems a group of people with smartphones cannot do more if more people with smartphones joint the group, unless they are also joined by a server, thus these systems do not fall under the definition of grassroots.
An example of a platform that is grassroots in spirit (even if not formally proven as such) is Scuttlebutt~\cite{tarr2019secure,kermarrec2020gossiping}.

\mypara{Motivation} 
Key grassroots platforms include grassroots social networks~\cite{shapiro2023gsn},  grassroots cryptocurrencies~\cite{shapiro2024gc,lewis2023grassroots}, and grassroots democratic federations~\cite{halpern2024federated}. Previously,  grassroots platforms were defined formally using unary distributed transition systems, in which each transition is carried out by a single agent~\cite{shapiro2023grassrootsBA}. 
However, grassroots platforms cater for a more abstract specification using transactions carried out atomically by multiple agents, something that cannot be expressed by unary transition systems.  
As a result, their original specifications were more cumbersome and opaque than they should be.
Moreover,  in grassroots platforms transactions carried out by different sets of participants need not be synchronized with each other, which greatly simplifies their mathematical foundations.

Here, we aim to provide a crisp formal foundation for these key grassroots platforms and beyond.  To do so, we enhance the notion of a distributed transition system to include atomic transactions and revisit the notions of grassroots platforms and their implementation within this enhanced foundation.  

Furthermore, previous work~\cite{shapiro2023grassrootsBA} provided sufficient conditions for when a platform (formally, a protocol defined via a family of distributed transition systems) is grassroots.  While going through the route presented in previous work is possible, the new foundations offer a simpler, more direct  and mathematically preferable way to prove that a platform, specified via atomic transactions, is grassroots.  Here, we follow this simpler path.

\mypara{Contributions} This paper:
\begin{enumerate}
    \item Presents atomic transactions and how they induce distributed transition systems.
    \item Provides crisp specifications of three key grassroots platforms via atomic transactions: Social networks, cryptocurrencies, and democratic federations.
    \item Provides a simpler and more stringent definition of grassroots platforms.
    \item Provides a sufficient condition for a transactions-based protocol to be grassroots.
    \item Shows that the atomic transactions employed in specifying each of the three key platforms satisfy this condition and therefore are are indeed grassroots.
\end{enumerate}

\mypara{Grassroots Social Networks} As a concrete example, consider grassroots social networks. Their goal is to provide people with the functionality of, say, Twitter/X-like feeds and WhatsApp-like groups, but without subjecting people, their social graph, or their personal information to the control of and exploitation by Musk, Zuckerberg, et al.  A key component in a social network is the social graph, encoding  who is a friend or a follower of whom.  In centralized social networks (e.g. Twitter/Facebook/WhatsApp) the social graph is stored, controlled, and commercially-exploited~\cite{zuboff2019age,zuboff2022surveillance} by the central authority.  In proposed decentralized architectures for social networks~\cite{DSNP}, the social graph is stored on the globally-replicated blockchain, under the control of a third-party---the miners/validators that execute the blockchain consensus protocol, who are remunerated for their service.

In a grassroots social network, the social graph is stored in a distributed way under the control of the people that participate in the network, with each person storing the local neighbourhood of the graph pertaining to them, and no third-party having access to any fragment of  the social graph unless explicitly granted. Here is a fragment of the specification of the social-graph maintenance protocol of a grassroots social networks.  
It assumes that each agent maintains, as its local state, a set of friends; that two agents $p$ and $q$ can atomically become friends, if they were not friends beforehand; and that the two friends $p$ and $q$ can atomically cease to be friends. 
It employs two binary atomic transactions:
\begin{tcolorbox}[colback=gray!5!white,colframe=black!75!black,top=2pt,bottom=2pt]
\begin{enumerate}
    \item \textbf{Befriend}:    $c'_p:= c_p\cup  \{q\}$,  
                                $c'_q:= c_q\cup  \{p\}$, provided $q\notin c_p$ and $p \notin c_q$
    \item \textbf{Unfriend}:    $c'_p:= c_p\setminus  \{q\}$,  
                                $c'_q:= c_q\setminus  \{p\}$, provided $q\in c_p$ and $p \in c_q$ 
\end{enumerate}
\end{tcolorbox}

Communication functions of a social network can be added, under the restriction  that communication occurs only among friends~\cite{shapiro2023gsn}.  For example, feeds with followers, where friends follow each other, and if two friends follow a third person, then the first to obtain an item on the third person's feed disseminates it to the other.
Formally, a liveness theorem can be proven for this design~\cite{shapiro2023gsn}, stating that if a person $p$ that follows a person $q$ is connected to $q$ via a chain of mutual friends, each of them correct and follows $q$, then $p$ will eventually receive every item on $q$'s feed.  
Practically, a true ``network celebrity'' employing this protocol may achieve efficient large-scale distribution of their feed via the friendship subgraph of their followers.

\mypara{Grassroots Cryptocurrencies}  
Grassroots cryptocurrencies~\cite{shapiro2024gc,lewis2023grassroots} can provide a foundation for an egalitarian digital economy, where people price their goods and services in terms of their own personally-minted grassroots coins, and liquidity is achieved through the creation of mutual credit lines among persons (natural and legal) via the swap of personal coins.  Grassroots coins issued by different persons form an integrated digital economy via the sole rule in the grassroots cryptocurrencies protocol:\footnote{This redemption rule is weaker than the one presented in previous work~\cite{shapiro2024gc}. The reason is that as long as $q$ holds $p$-coins it can dodge a redemption claim by $p$ by issuing a counter-claim, until the mutual credit line between $p$ and $q$ dries up.  Following that, $p$ can force redemption only if, to begin with, $p$ had a positive credit balance with $q$, namely held more $q$-coins than the number of $p$-coins held by $q$.  This weaker redemption obligation  eschews such `redemption wars':   It  allows $p$ to redeem $q$-coins it holds against non-$p$-coins held by $q$ when their credit line ceases to be mutual, while of course not excluding voluntary swaps by mutual consent at any time, which we expect to be the norm.}
\begin{tcolorbox}[colback=gray!5!white,colframe=black!75!black,top=2pt,bottom=2pt]
\textbf{Coin Redemption:} $p$ can redeem a $q$-coin it holds against a $p$-coin held by $q$, if there is one, else against any coin held by $q$, one-for-one.
\end{tcolorbox}
We note that the value of $p$-coins critically depends on $p$ maintaining computational and economic integrity.  In the specification of grassroots cryptocurrencies each agent maintains, as its local state, the set of coins they hold; each agent $p$ may mint additional $p$-coins  (¢$_p^k$ denotes $k$ $p$-coins); and two agents $p$ and $q$ can atomically swap any coins they hold.  
The following fragment of the specification employs one unary transaction and one binary atomic transaction:
\begin{tcolorbox}[colback=gray!5!white,colframe=black!75!black,top=2pt,bottom=2pt]
\begin{enumerate}
    \item \textbf{Mint}: $c'_p := c_p \cup $¢$_p^k$, $k>0$.
    \item \textbf{Swap}:    $c'_p:= (c_p\cup  y) \setminus  x$, 
                                $c'_q:= (c_q\cup  x) \setminus  y$,
                                provided
                                $x \subseteq c_p$ and  $y \subseteq c_q$.
\end{enumerate}
\end{tcolorbox}

The swap transaction can realize the key economic functions of grassroots cryptocurrencies:
\begin{enumerate}
    \item \textbf{Payment}:  Paying $q$ with $q$-coins, namely  $x = $¢$_q^j$, $j>0$, and $y=\emptyset$ (also paying with other coins $q$ accepts as payment, e.g. community-bank coins). A payment could be made for love or, more typically, expecting in exchange (or after having received) an `off-chain' product or service from $q$.  In a practical application $y$ could include a payment receipt, a confirmed purchase order, etc.
    
    \item \textbf{Mutual Credit Lines}:  Establish mutual credit lines via the swap of personal (self-minted) coins, namely  $x = $¢$_p^j$ and $y=$¢$_q^k$ for some $j,k>0$.  Typically $j=k$, but mutual credit with a premium for one of the agents, namely with $j\ne k$, is also possible. A digital economy based on grassroots cryptocurrencies achieves liquidity through persons (natural and legal) establishing mutual credit lines among them.
      
    \item \textbf{Redemption}: The obligatory 1:1 swap of $q$-coins held by $p$ against an equal number of coins held by $q$, namely $x=$¢$_q^j$, $|y|=j$.
\end{enumerate}

\mypara{Grassroots Democratic Federation}
Grassroots Democratic Federation aims to address the democratic governance of large-scale decentralized digital communities, e.g., of the size of the EU, the US, existing social networks, and even humanity at large.
A grassroots democratic federation evolves via the grassroots formation and federation of digital communities, each governed by an assembly selected by sortition from its members. The approach assumes digital freedom of assembly---that people can freely form  digital communities that can federate into ever-larger communities as well as spawn child communities,  based on geography, relations, interests, or causes.
Small communities (say $<100$) are governed by their members.  Larger communities---no matter how larger---are each governed democratically by a small  (say $\approx 100$) assembly elected by sortition among its members. 

The challenge of specifying grassroots federation via atomic transactions is similar to the social network one, with a technical difference and a fundamental difference.  The technical difference is that in a grassroots social network the friendship graph is undirected, but the grassroots federations the graph is directed.  Furthermore, a federation is intended to offer a hierarchical democratic governance structure, therefore its directed graph must be acyclic. 

The fundamental difference between grassroots social networks and grassroots federations is that the actors in the former are individuals, whereas the actors in the latter are communities of individuals, governed by assemblies.  The standard centralized realization of this requirement would be to run a program on a server/cloud for each community, which realizes some democratic governance protocol by the assembly of that community.  The standard decentralized realization of this requirement would be to similarly run a \emph{smart contract}~\cite{de2021smart}, realizing a Decentralized Autonomous Organization (DAO)~\cite{ethereum:dao} for each community on some decentralized global platform that providers a consensus protocol needed to execute smart contracts.  Neither is grassroots.  
The grassroots solution would be to realize each community as a \emph{digital social contract}~\cite{cardelli2020digital}, which also requires a grassroots consensus protocol among the participants, namely one that is executed by the participants themselves, not by third-parties (miners, validators) operating a global platform.

A specification of grassroots federation via unary distributed transition systems has yet to be attempted. It would require a grassroots consensus protocol by which members of two communities decide, atomically, to join or leave.  Here, we define such atomic transactions abstractly:  All members of a community carry its state locally; and all of them change its state  atomically to realize joining or leaving another community.  The details are shown below.

\mypara{Paper outline} Section \ref{section:dts} introduces distributed transition systems and how they are induced by a set of atomic transactions.  Section \ref{section:gs-platforms} provides transactions-based specifications of three key grassroots platforms:  Grassroots social networks, grassroots cryptocurrencies, and grassroots democratic federations. Section \ref{section:gs-protocols} introduce protocols, grassroots protocols, protocols induced by atomic transactions, interactive atomic transactions, and proves the main theorem: That a protocol induced by an interactive set of transactions is grassroots.  It then shows that the three key platforms introduced here are interactive, and concludes that they are grassroots as specified.  Section \ref{section:implementation} discusses, informally, the implementation of grassroots platforms.  Section \ref{section:conclusion} discussed related and future work.  

\section{Distributed Transition Systems and Grassroots Protocols}\label{section:dts}


In this section we define transition systems, distributed transition systems, atomic transactions, and grassroots protocols.

\subsection{Transition systems}\label{subsection:ts1}

The following is a simplified variation, sufficient for the purpose of this work, on the foundations introduced in~\cite{shapiro2021multiagent}.
In the following, $a \ne b \in X$ is a shorthand for $a\ne b \wedge  a \in X \wedge b \in X $.   

\begin{definition}[Transition system, Computation, Run]\label{definition:ts}
A \temph{transition system} is a tuple $TS=(S,s0,T)$, where: 
\begin{enumerate}
    \item $S$ is an arbitrary non-empty set, referred to as the set of \temph{states}.
    \item Some $s0\in S$ is called the \temph{initial state}. 
    \item $T\subset S^2$ is a set of \temph{transitions over} $S$, where each transition $t\in T$ is a pair $(s,s')$ of non-identical states $s\ne s'\in S$,  also written as $t=s\rightarrow s'$.
\end{enumerate} 
A \temph{computation} of $TS$ is a (nonempty, potentially infinite) sequence of states $r= s_1,s_2,\cdots$ such that for every two consecutive states $s_i,s_{i+1} \in r$, $s_i\rightarrow s_{i+1} \in T$.  If $s_1=s0$ then the computation is called a \temph{run} of $TS$.
\end{definition}
Given a computation $r= s_1,s_2,\ldots$, we use $r\subseteq T$ to mean $(s_i\rightarrow s_{i+1}) \in T$ for every $(s_i\rightarrow s_{i+1}) \in r$.

We note that the definition of transition systems originally employed in the definition of grassroots systems~\cite{shapiro2021multiagent, shapiro2023grassrootsBA} included a liveness condition.  As the rather-abstract atomic transactions employed here are volitional and have no associated liveness conditions, these, are omitted to simplify the exposition.
Also, the original work considered faulty computations and fault-tolerant implementations, not considered here.  The  simplified-away components could be re-introduced in follow-on work that considers live and fault-tolerance implementations of the specification
presented here.

\subsection{Distributed Transition Systems with Atomic Transactions}\label{subsection:mts}

We assume a potentially infinite set of \emph{agents} $\Pi$ (think of all the agents that are yet to be born), but consider only finite subsets of it,  so when we refer to a particular set of agents $P \subset \Pi$ we assume $P$ to be nonempty and finite.  We use $\subset$ to denote the strict subset relation and $\subseteq$ when equality is also possible.

In the context of distributed systems it is common to refer to the state of the system  as \emph{configuration}, so as not to confuse it with the \emph{local states} of the agents.
As standard, we use $S^P$ to denote the set $S$ indexed by the set $P$, and if $c\in S^P$ we use $c_p$ to denote the member of $c$ indexed by $p\in P$.  Intuitively, think of such a $c\in S^P$ as an array of cells indexed by members of $P$ and with cell values in $S$.


\begin{definition}[Local States, Configuration, Transitions, Active \& Stationary Participants, Degree]
Given agents $P \subset \Pi$ and an arbitrary set $S$ of \temph{local states} with a designated \temph{initial local state} $s0\in S$, a \temph{configuration} over $P$ and $S$ is a member of $C:= S^P$ and the \temph{initial configuration} is $c0:= \{s0\}^P$.  A \temph{distributed transition} over $P$ and $S$ is any pair of configurations $t=c\rightarrow c' \in C^2$ s.t. $c\ne c'$, with $t_p := c_p \rightarrow c'_p$ for any $p\in P$, and with $p$ being an \temph{active} in $t$ if $c_p\ne c'_p$, \temph{stationary participant} otherwise.  The \temph{degree} of $t$ (unary, binary,\ldots $k$-ary)
is the number of active participants in $t$.
A unary transition with active participant $p$ is a \temph{$p$-transition}, and the degree of a set of transitions is the maximal degree of any of its members.
\end{definition}

\begin{definition}[Distributed Transition System]\label{definition:dts-cd}
Given agents $P \subset \Pi$ and an arbitrary set $S$ of \temph{local states} with a designated \temph{initial local state} $s0\in S$, 
a \temph{distributed transition system} over $P$ and $S$ is a transition system $TS= (C,c0,T)$ with $C:= S^P$,  $c0:= \{s0\}^P$, and $T\subseteq C^2$ being a set of distributed transitions over $P$ and $S$.
\end{definition}
Unary distributed transition systems were introduced in~\cite{shapiro2021multiagent} and were employed to define the notion of grassroots protocols~\cite{shapiro2023grassrootsBA} and to provide unary specifications for various grassroots platforms~\cite{shapiro2023gsn,shapiro2024gc,lewis2023grassroots}.
Here, we employ $k$-ary transition systems, for any $k\le |P|$, in which several agents can change their state simultaneously.
However, rather than specifying a transition system directly, we specify it via the
atomic transactions~\cite{lynch1988theory} the transition system intends to realize.

Informally, an atomic transaction:
\begin{enumerate}
    \item Can have several active participants (be binary, $k$-ary) that change their local states atomically as specified. 
    \item Specifies explicitly its participants (both active and stationary),  thus implicitly defining infinitely-many distributed transitions where all non-participants remain in their arbitrary state.
\end{enumerate}
In addition to being atomic, the description above implies that transactions are asynchronous~\cite{shapiro2021multiagent}, in the sense that if a transaction can be carried out by its participants, with non-participants being in any arbitrary states, it can still be carried out no matter what the non-participants do. In particular, the active participants in a transaction need not synchronize its execution with any non-participant (but they must synchronize, of course, with stationary participants, as changing the local state of any participant would disable the transaction).  Informally, a transaction is but a distributed transition over its participants, which defines a set of distributed transitions over any larger set of agents.
Formally:
\begin{definition}[Transaction, Closure]\label{definition:closure}
Let $P\subset \Pi$, $S$ a set of local states, and $C:=S^P$.
A \temph{transaction} $t=(c\rightarrow c')$ over local states $S$ with \temph{participants} $Q \subset \Pi$ is but a distributed transition over $S$ and $Q$.
For every $P\subset \Pi$ s.t.  $Q\subseteq P$, the \temph{$P$-closure of $t$}, $t{\uparrow}P$,  is the set of transitions over $P$ and $S$ defined by:
$$
t{\uparrow}P := \{ t' \in C^2  :
\forall p\in Q.(t_p = t'_p) \wedge \forall p\in P\setminus Q.(p\text{ is stationary in }t')\}
$$
If $R$ is a set of transactions, each $t\in R$ over some $Q\subseteq P$ and $S$, then the 
\temph{$P$-closure of $R$}, $R{\uparrow}P$, is the set of transitions over $P$ and $S$ defined by:
$$
R{\uparrow}P := \bigcup_{t\in R} t{\uparrow}P
$$
\end{definition}
Namely, the closure over $P\supseteq Q$ of a transaction $t$ over $Q$ includes all transitions $t'$ over $P$ in which members of $Q$ do the same in $t$ and in $t'$, and the rest remain in their current (arbitrary) state.

Note that while the set of distributed transitions $R{\uparrow}P$ in general may be over a larger set of agents than $R$, $T$ and $R$ are of the same degree, by construction.
Also, note that while each of the distributed transitions of a distributed transition system over $P$ is also over $P$,  in a set of transactions each member is typically over a different set of participants.  Thus, a set of transactions $R$ over $S$, each with participants in some $Q\subseteq P$, defines a distributed transition system over $S$ and $P$ as follows:

\begin{tcolorbox}[colback=gray!5!white,colframe=black!75!black,top=4pt,bottom=4pt]
\begin{definition}[Transactions-Based Distributed Transition System]\label{definition:tbdts}
Given agents $P \subset \Pi$, local states $S$ with initial local state $s0\in S$, 
and a set of transactions $R$, each $t\in R$ over some $Q\subseteq P$ and  $S$, a \temph{transactions-based distributed transition system} over $P$, $S$, and $R$ is the distributed transition system $TS= (S^P,\{s0\}^P,R{\uparrow}P)$ .
\end{definition}
\end{tcolorbox}
In other words, one can fully specify a distributed transition system over $S$ and $P$ simply by providing a set of atomic transactions over $S$, each with participants $Q\subseteq P$.  This is what we do next for grassroots social networks, grassroots cryptocurrencies, and grassroots democratic federations.

\section{Grassroots Platforms with Atomic Transactions}\label{section:gs-platforms} 

Grassroots protocols were defined formally using unary distributed transition systems.  Thus, higher-level protocols that include atomic transactions by multiple agents could not be expressed directly within this formalism, let alone proven grassroots.

Yet, the grassroots platforms we are interested in---grassroots social networks, grassroots cryptocurrencies, and grassroots democratic federations---all call for atomic transactions for their specification.   Therefore, we extended above the notion of a distributed transition system to include atomic transactions carried out by a set of agents. In this section we revisit the grassroots platforms previously defined using unary transition systems and redefine them more simply and abstractly using binary atomic transactions.  In addition, we present a $k$-ary transactions-based specification of grassroots democratic federations, which where hitherto defined only at the community level.

\subsection{Grassroots Social Network via Befriending and Unfriending}

A grassroots social network evolves by people forming and breaking friendships.  The original definition of grassroots social networks~\cite{shapiro2023gsn} was via a unary distributed transition system.  Here both actions are specified---abstractly and concisely---as binary transactions.
The local state of an agent $p\in P$ is the set of agents $Q\subseteq P$ of which $p$ is a friend.

\begin{definition}[Grassroots Social Network]
We assume a given set of agents $P\subset \Pi$.
The \temph{grassroots social network transition system} over $P$ is the distributed transition system $SN$ over $P$, $S := 2^{P}$, and the binary transactions $c\rightarrow c'$ with every  $\{p,q\}\subseteq P$ as participants, satisfying:
\begin{enumerate}
    \item \textbf{Befriend}:    $c'_p:= c_p\cup  \{q\}$,  
                                $c'_q:= c_q\cup  \{p\}$, provided $q\notin c_p$ and $p \notin c_q$
    \item \textbf{Unfriend}:    $c'_p:= c_p\setminus  \{q\}$,  
                                $c'_q:= c_q\setminus  \{p\}$, provided $q\in c_p$ and $p \in c_q$
                                
\end{enumerate}
\end{definition}

\begin{lemma}[Friendship Safety]\label{lemma:friendship-safety}
Given a run $r$ of $SN$, a configuration $c\in r$, and agents $p, q\in P$,
then $p\in c_q$ iff $q\in c_p$.
\end{lemma}
\begin{proof}
The proof is by induction on the length of the run $r=c0,c_1,\ldots,c_n$.  Assume $|r|=1$, namely $r$ consists of the initial configuration $c0$.  Then all local states are empty, satisfying the Lemma.
Assume the lemma holds for runs of length $|r|=n$, that $c_n$ satisfies the Lemma, and consider the transition $c_n\rightarrow c_{n+1}$.  It can be either Befriend or Unfriend; both satisfy for every $p\in P$ the condition $p\in c_q$ iff $q\in c_p$ for $c_{n+1}$ if $c_n$ does.
\end{proof}

We note that each configuration $c$ in a run $r$ of $GC$ induces a graph with agents as vertices and an edge $p\leftrightarrow q$ if $p\in c_q$ and $q\in c_p$, and that the graphs induced by two consecutive configurations in $r$ differ by exactly one added or removed edge.

\subsection{Grassroots Cryptocurrencies via Atomic Swaps}

The original specification of grassroots cryptocurrencies was via unary distributed transition system and hence was quite involved, and so was the proof of them being grassroots~\cite{shapiro2024gc}. The definition in turn led to an implementation via a unary payment system~\cite{lewis2023grassroots}. Here we provide an alternative specification---abstract and concise---via binary transactions.  We will later prove the specification to be grassroots.

Informally, we associate with each agent $p\in P$ a unique `colour' and an infinite multiset of identical $p$-coloured coins.  Formally:
\begin{definition}[Coins]\label{definition:coins}
We assume an infinite multiset $C$ of identical \temph{coins} ¢. Given set of agents $P\subset \Pi$, we refer to ¢$_p$, a $p$-indexed element of the indexed set $C^P$, as a $p$-coin, with  ¢$_p^k$ being
a multiset of $k$ $p$-coins.
A \temph{set of $P$-coins} is a member of $2^{C^P}$, namely a multiset of $P$-indexed coins.
\end{definition}

The transactions of grassroots cryptocurrencies include the unary minting of $p$-coins by $p$, and the atomic swap between $p$ and $q$ of a set of coins held by $p$ in exchange for a set of coins held by $q$.

\begin{definition}[Grassroots Cryptocurrencies]\label{definition:gc}

The \temph{grassroots cryptocurrencies transition system} over $P$ is the distributed transition system $GC$ over $P$,  $S := 2^{C^P}$, and transactions $c\rightarrow c'$ with every  $\{p,q\}\subseteq P$ as participants, satisfying: 
\begin{enumerate}
    \item \textbf{Mint}: $c'_p := c_p \cup $¢$_p^k$, $k>0$.
    \item \textbf{Swap}:    $c'_p:= (c_p\cup  y) \setminus  x$, 
                                $c'_q:= (c_q\cup  x) \setminus  y$,
                                provided
                                $x \subseteq c_p$ and  $y \subseteq c_q$.
\end{enumerate}
\end{definition}

The swap transaction can realize several different economic functions, including payments, and opening of mutual credit lines, and coin redemption,  as described in the Introduction.

The present specification does not distinguish between voluntary swap transactions,  namely payments and forming mutual credit, and obligatory swaps, namely coin redemption.  
This issue is further discussed below.

\begin{lemma}[Safety: Conservation of Money]\label{lemma:conservation}
Given a run $r$ of $GC$, a configuration $c\in r$ and an agent $p\in P$,
the number of $p$-coins in $c$ is the the number of $p$-coins minted by $p$ in the prefix of $r$ ending in $c$.
\end{lemma}
\begin{proof}
The proof is by induction on the length of the run $r=c0,c_1,\ldots,c_n$.  Assume $|r|=1$, namely $r$ consists of the initial configuration $c0$.  Then all local states are empty, satisfying the Lemma.
Assume the lemma holds for runs of length $|r|=n$, that $c_n$ satisfies the Lemma, and consider the transition $c_n\rightarrow c_{n+1}$.  If the transition is Mint and the claim holds for all, then after minting it still holds by adding the minted coin.  If it is Swap, it still holds as Swap does not adds new coins to the configuration or removes coins from it.
\end{proof}

\subsection{Grassroots Democratic Federation via Joining and Leaving}

\mypara{Background} Grassroots federation~\cite{halpern2024federated} is a process by which communities evolve and  federate with other communities.  Key steps include  a community joining a federation, by the mutual consent of the two, and a community leaving a federation, by the unilateral decision of one of the two.  

In the following $G=(V,E)$ is a federation graph with communities as nodes and directed edges indicating community relation,  specifically $f\rightarrow v$ indicates that $v$ is a  child community of $f$, and $G'$ is the graph resulting from applying any of the following transitions to $G$:
\begin{enumerate}

\item \textbf{Federate $v\in V$:} Add a new node $f\notin V$ to $V$ and the edge  $f\rightarrow v$ to $E$.
 
    \item \textbf{Join $f\rightarrow v$:} Add $f\rightarrow v$ to $E$ provided that $f\in V$ and  $G'$ is acyclic.

    \item \textbf{Leave $f \rightarrow v$:}  If  $f\rightarrow v \in E$ then remove  $f\rightarrow v$ from $E$.
\end{enumerate}

\mypara{Challenges} As discussed in the Introduction, the challenge of specifying grassroots federation via atomic transactions is similar to the social network one, with a technical difference and a fundamental difference.  The technical difference is that in a grassroots social network the friendship graph is undirected, but the grassroots federations the graph is directed.  Furthermore, a federation is intended to offer a hierarchical democratic governance structure, therefore its directed graph must be acyclic. 
To allow local transactions to evolve the federation without creating cycles, a partial order $\succ$ on communities can be devised, with conditioning the joining of $g$ as a child of $f$ on  $f\succ g$. 
Here are several examples for such a partial order:
\begin{enumerate}
    \item \textbf{Height:} Let $h(f)$ be the maximal length of any path  from $f$.  Then $f\succ g$ if $h(f) > h(g)$.
    \item \textbf{Generality:}  Assume some topic tree, e.g. with nodes being animal lovers, dog lovers, husky lovers, malamute lovers with the obvious edges, and associate a topic with each federation.  Then $f\succ g$ if there is a nonempty path in the topic tree from the topic of $f$ to the topic of $g$.
    \item \textbf{Geography:}  Associate with each federation a region on the globe. Then $f\succ g$ if the region of $f$  strictly includes the region of $g$.
\end{enumerate}
In a real deployment of grassroots federations we expect the partial order to be a superposition of geography, topics, and maybe other issues,  so that the federation of dog lovers of a village can join both the general village federation and the regional federation of dog lovers, which in turn can join the general regional federation.  The resulting federation structure, termed \emph{laminar}, was investigated in~\cite{halpern2024federated}.  With this constraint we can address the requirement that the federation graph always be acyclic.

A related technical issue that arises in a grassroots setting is to ensure, without central coordination, that each community has a globally-unique identifier. This is required so that a grassroots federation can scale indefinitely, integrating communities that have emerged independently of each other.  To achieve that we assume that each agent chooses for itself a unique key-pair at random, without collisions, and we  identify each agent with its public key.   Any agent $p$ may create one singleton community with itself as a member, identified by $p$.  A community with a unique identifier $c$ may create any number of parent communities using the \textbf{Federate} transition, and endows the $i^{th}$ parent community it creates with the unique identifier $(c,i)$. 

Here, we employ community identifiers $P^*$, where each $c\in P^*$ consists of an agent $p\in P$ followed by a finite list of integers, and define a total (not partial) order $\succ$ on  $P^*$,  where $v\succ v'$  for $v,v'\in P^*$ if the list $v$ is longer than the list $v'$, with ties resolved lexicographically.

The fundamental difference between grassroots social networks and grassroots federations is that the actors in the former are individuals, whereas the actors in the latter are communities of individuals.  
We address it as follows:
All members of a community hold an identical copy of its state, and a transaction between two communities, say $g$ joining $f$, is realized by the corresponding $k$-ary atomic transaction among the members of the two communities, with $k$ being the size of their union, 
in which the states of all members of both communities  change atomically to reflect this federation-level transaction.

Naturally, this results in many aspects of the transaction being abstracted-away, including that:
\begin{enumerate}
    \item Joining requires mutual consent and leaving is unilateral.
    \item Decisions on behalf of a community are taken by the assembly of a community.
    \item The decision process is constitutional and democratic.
\end{enumerate}

\mypara{Specification} A \emph{federation graph} $G=(V,E)$ over $P$ is a labelled directed acyclic graph with nodes labelled by community identifiers; we do not distinguish between nodes and their labels.
The \emph{initial federation graph} over $P$  is $G_0(P)=(P,\emptyset)$.  Let $\calG(P)$ be the set of federation graphs over $P$.

\begin{definition}[Community and Personal Subgraph]
Given a federation graph $G=(V,E\in \calG(P)$,  
For a community $v\in V$, the \temph{community subgraph} $G_v=(V_v,E_v)$ of $v$ is the subgraph of $G$ where $V_v$ includes $v$ and all nodes adjacent to $v$ and $E_v$ includes all edges incident with $v$.
An agent $p\in P$ is a \temph{member} of community $v\in G$ if there is a path in $G$ from $v$ to $p$, and the \temph{personal subgraph} $G_p$ of $p$ is the union of the subgraphs of the communities $p$ is a member of, namely $G_p := \bigcup_{p\in v} G_v$
\end{definition}
Thus, $G_p$ includes any community $v$ that $p$ is a member of and any edge incident with $v$

\begin{observation}\label{observation:G}
Let $G\in \calG(P)$ for some set of agents $P\subset \Pi$.
Then $G = \bigcup_{p\in P} G_p$.
\end{observation}
\begin{proof}[Proof of Observation \ref{observation:G}]
Let $G=(V,E)$ as above. We prove the two directions of equality:
\begin{enumerate}
    \item $\subseteq$:  Let $v\in V$.  Then some $p\in P$ is a member of $v$ by construction.  Hence $v\in V_p$.  Let $f\rightarrow g\in E$, with some $p\in V_f$.  Then by definition $p\in V_g$, and hence $f\rightarrow g \in E_p$.
    \item $\supseteq$ Let $p\in P$ and $v\in V_p$.  Then $p$ is a member of $v$ in $G$, and therefore $v\in V$.  Let  $e\in E_p$.  By construction, this can be the case only if $e\in E$.
\end{enumerate}
\end{proof}

Namely, a federation graph is fully-defined by its members' subgraphs. Hence, in the following grassroots federation transition system, the evolving federation graph $G$ created during a run is stored in a distributed way, with each agent $p\in P$ maintaining $G_p$.
The result is that each agent stores the state of all communities they are a member of; equivalently, the state of each community is stored by all its members.  Formally:

\begin{definition}[Federation Configuration, Valid]
A \temph{federation configuration} $c$ over $P\subset \Pi$ has local states 
$\{G_p: G \in \calG(P), p\in P\}$.  The \temph{initial federation} is $G0:=(V0,\emptyset) \in \calG(P)$, where $V0$ has a $\{p\}$-labelled node for every $p\in P$, and the \temph{initial federation configuration} $c0$ is defined by $c0_p  :=  G0_p:= (\{p\},\emptyset)$.
A federation configuration is \temph{valid} if there is a graph $G\in \calG(P)$ such that $c_p = G_p$ for every $p\in P$.
\end{definition}
Note that the initial federation configuration is valid.

The atomic transactions of grassroots federation ensure that when the state of a community $v$ changes in a graph $G$ through the addition or removal of an edge incident to $v$ in $G$,  this change is carried out atomically by all $p\in P_v$, each updating its local state $G_p$.

This formulation can be considered as an abstraction of direct democracy, as every action of a community is carried out by all its members.  This would mean that realizing this model requires all members of a community to jointly run a consensus protocol, the results of which define the acts of the community.  While simple to specify, this approach may be unrealistic for large communities.  For this and other reasons, works on grassroots federation~\cite{halpern2024federated} explore a model in which each community is governed by a small (say, ${\approx}100$) assembly, selected by sortition among the members of the community.  In such a model, only assembly members would store the state of the community, and only the assembly, not the entire community, would run the consensus protocol realizing its behaviour.  It is important to note that in both formulations, the consensus (between the entire population, or just the assembly) is on the decision taken.  In case of an assembly, the decision process itself could be ``on-chain'', e.g. implemented by a social contract~\cite{cardelli2020digital} realizing the various methods developed for the egalitarian democratic formation and governance of digital communities~\cite{poupko2021building,shahaf2020genuine,shahaf2018sybil,meir2024safe,abramowitz2021beginning,bulteau2021aggregation,elkind2021united,abramowitz2021democratic}, or ``of-chain'', e.g. via an online video meeting of the members of the assembly in which votes are taken by show of hands.  The only requirement is that there will be consensus among the members of the assembly on the democratic decision that was taken.
Here, we stick with the direct-democracy model to simplify the exposition and are agnostic/abstract-away the decision process employed.

\begin{definition}[Grassroots Federation]
A \temph{grassroots federation transition system} over $P\subseteq \Pi$ is a distributed transition system $GF$ over $P$ and $\calG(P)$,  with every transaction $c\rightarrow c'$ for every $G=(V,E)\in \calG(P)$  such that:
\begin{enumerate}
    \item \textbf{Federate $v$}: $c, c'$ are configurations over $Q:=P_v$, $v\in V$, $f=(v,i+1)$, where $i$ is the maximal index of a parent of $v\in G$,  if any, $i=0$ if none,  and\\ 
    $\forall p\in Q.(c_p = G_p \wedge c'_p := (V \cup  \{f\}, E\cup \{f\rightarrow v\})_p)$.
 
   \item \textbf{Join $f\rightarrow g$}:  $c, c'$ are configurations over $Q:=P_f\cup P_g$, $f,g\in V$, $f\succ g$,  and\\
   $\forall p\in Q.(c_p=G_p \wedge  c'_p := (V, E \cup \{f\rightarrow g\})_p)$.

   \item \textbf{Leave $f\rightarrow g$}:  $c, c'$ are configurations over $Q:=P_f$, $f,g\in G$, $f\rightarrow g \in E$,  and\\
   $\forall p\in Q.(c_p=G_p \wedge  c'_p := (V, E \setminus \{f\rightarrow v\})_p)$.
\end{enumerate}
\end{definition}

\begin{lemma}[Federation Safety]
In a run $r$ of $GF$, any configuration $c\in r$ is valid.
\end{lemma}
\begin{proof}
The proof is by induction on the length of the run.
Initially, $G = G0$ and the initial configuration $c0$ satisfies $G0=\bigcup_{p\in P} c0_p$.
Assume that configuration $c$ is valid and encodes $G=(V,E)$,  and consider a transition $c\rightarrow c'$ with $c'$ encoding $GF'$, namely $G'=\bigcup_{p\in P} c'_p$.  We consider the various cases:
\begin{enumerate}
 \item \textbf{Federate $v$}: Each $c\in P_v$ changes their state to $c'_p = (V \cup  \{f\}, E\cup \{f\rightarrow v\})_p)$.
 
   \item \textbf{Join $f\rightarrow g$}: Each  $p\in (P_f \cup P_g)$ changes their state to $c'_p=(V, E \cup \{f\rightarrow g\})_p)$.

   \item \textbf{Leave $f\rightarrow g$}:  Each $p\in P_f$, which includes $P_g$,  changes their state to  $c'_p= (V, E \setminus \{f\rightarrow v\})_p)$.
\end{enumerate}
Examining these changes shows that they all preserve validity, in that also in the new configuration each agent $p$ records, by construction, exactly the $p$-projection of the updated federation graph $G'$.
\end{proof}

This completes the transactions-based specification of the three key grassroots platforms. 
It does not distinguish between transactions carried out by mutual consent (befriend, open a credit line, join), and  transactions that are obligatory once requested (unfriend, redeem, leave).
The distinction can be added at this level of abstraction, with a liveness requirement that once a request to carry out an obligatory transaction is made, it is eventually carried out.  Alternatively, it can be added at the unary implementation level; this may be more natural as atomic transactions would anyway be realised by one agent issuing a request/offer and the other(s) consenting to it~\cite{shapiro2023gsn,lewis2023grassroots}.

\section{Grassroots Protocols}\label{section:gs-protocols}

Here we define what is a protocol; 
define when a protocol is grassroots; 
show how to define a protocol via a set of transactions; 
present conditions under which a protocol defined via transactions is grassroots;  
argue that the three protocols defined  via transactions above satisfy these conditions; 
and conclude that these three protocols are grassroots.

\subsection{Protocols and Grassroots Protocols}

A protocol is a family of distributed transition systems, one for each set of agents $P\subset \Pi$, which share an underlying set of local states $\calS$  with a designated initial state $s0$.
A \emph{local-states function} $S: P \mapsto 2^\calS$ maps every set of agents $P \subset \Pi$ to an arbitrary set of local states $S(P)\subset \calS$ that includes $s0$ and satisfies  $P\subset P' \subset \Pi \implies S(P) \subset S(P')$.  Given a local-states function $S$, we use $C$ to denote configurations over $S$, with $C(P) := S(P)^P$ and $c0(P):= \{s0\}^P$.



Next, we define the notions of a protocol and a grassroots protocol.
\begin{definition}[Protocol]\label{definition:family}
A \temph{protocol} $\calF$ over a local-states function $S$ is a family of distributed transition systems that has exactly one transition system $\calF(P) = (C(P),c0(P),T(P))$ for every $P \subset \Pi$.
\end{definition}


Informally, in a grassroots protocol a set of agents $P$, if embedded within a larger set $P'$, can still behave as if it is on its own, but has additional behaviours at its disposal due to possible interactions with members of $P'\setminus P$.
To define the notion formally we employ the notion of projection.

\begin{definition}[Projection]
Let $\emptyset \subset P \subset P' \subset \Pi$.
If $c'$ is a configuration over $P'$ then $c'/P$, the \temph{projection of $c'$ over $P$}, is the configuration $c$ over $P$ defined by $c_p := c'_p$ for every $p\in P$.  
%
\end{definition}
Note that in the definition above, $c_p$, the state of $p$ in $c$,  is in $S(P')$, not in $S(P)$, and hence may include elements ``alien'' to $P$, e.g.,  friendship with or a coin of $q\in P'\setminus P$.


We use the notions of projection and closure (Definition \ref{definition:closure}) to define the key notion of this paper, a grassroots protocol:
\begin{tcolorbox}[colback=gray!5!white,colframe=black!75!black,top=4pt,bottom=4pt]
\begin{definition}[Oblivious, Interactive, Grassroots]\label{definition:grassroots}
A  protocol $\calF$ is:
\begin{enumerate}
    \item \temph{oblivious} if for every $\emptyset \subset P \subset P' \subseteq \Pi$, 
    $T(P){\uparrow}P'\subseteq T(P')$
    \item  \temph{interactive} if for every $\emptyset \subset P \subset P' \subseteq \Pi$ and every configuration $c\in C(P')$ such that  $c{/}P\in C(P)$, there is a computation  $c\xrightarrow{*} c'$ of $\calF(P')$ for which $c'{/}P\notin C(P)$.
    \item \temph{grassroots} if it is oblivious and interactive.
\end{enumerate}
\end{definition}
\end{tcolorbox}
\mypara{Oblivious} Being oblivious implies that if a run of $\calF(P')$ reaches some configuration $c'$, then anything $P$ could do on their own in the configuration $c'/P$ (with a transition from $T(P))$, they can still do in the larger configuration $c'$ (with a transition from $T(P')$), effectively being oblivious to members of $P'\setminus P$.
The reason is that if $t=(c'/P\rightarrow d)\in T(P)$, then by the definition of closure,  $t'=(c'\rightarrow d') \in t{\uparrow}P'\subseteq T(P')$, where $d'_p = d_p$ if $p\in P$ else $d'_p= c'_p$.   Inductively, this implies that if agents $P$ employ only transitions in $T(P)$ from the start, with their local states remaining in $S(P)$, they could continue doing so indefinitely, effectively ignoring any members in $P'\setminus P$.
Proving the corresponding claim within the original definition was more difficult, and required the notion of interleaving of computations of two distributed transition systems~\cite{shapiro2021multiagent}.

We note that any protocol that uses a global data structure, whether replicated (Blockchain) or distributed (DHT~\cite{rhea2005opendht}, IPFS~\cite{benet2014ipfs}), is not oblivious as members of $P$ cannot ignore changes to the global data structure made by members of $P'\setminus P$; and hence is not grassroots.

\mypara{Interactive} A slight complication in the definition of interactive is the use of a finite computation $\xrightarrow{*}$ rather than just a single transition $\rightarrow$.  It is required  as in some grassroots platforms agents need to make some preparatory steps before they may interact with each other.  For example, in grassroots cryptocurrencies agents need first to mint some coins before they can swap them. 

Being interactive is a weak liveness requirement.  Informally,  a standard liveness requirement has the form  ``something good must eventually happen''.  Here, the requirement is ``it must be the case that something good ($P$ interacting with non-$P$) may eventually happen''. Namely, no matter what members of $P$ do, it is always the case that they can eventually interact with non-$P$'s.  Moreover, being interactive not only requires that $P$ can always eventually interact with $P'\setminus P$, but that they do so in a way that leaves ``alien traces'' in the local states of $P$, so that the resulting configuration $c'/P$ could not have been produced by $P$ running on their own.  
For example, forming friendships with,  or receiving coins from, agents outside of $P$.
We note that while federated systems~\cite{raman2019challenges,bittorrent} are oblivious, as one server may choose to ignore all other servers and just serve its clients, they are not interactive, as a  set of clients $P$ without a server cannot do more if embedded within a larger set of clients $P'$, still without a server.



\subsection{Transactions-Based Grassroots Protocols}

The original paper~\cite{shapiro2023grassrootsBA} provided sufficient conditions for a protocol to be grassroots---asynchrony, interference-freedom, and interactivity.  They should still hold under the new definition, with some adaptation, mostly simplification.  However, here we are interested in transactions-based transition systems, therefore we will follow a different route:  We first show how a local-states function can be used to define a set of transactions, and then show how a set of such transactions can be used to define a protocol, termed transactions-based protocol.  We then prove that a single condition on such transactions---interactivity---is sufficient for a transactions-based protocol to be grassroots.

\begin{definition}[Transactions Based on a Local-State Function]\label{definition:tblsf}
Let $S$ be a local-states function.
A set of transactions $R$ is \temph{over $S$} if every transaction $t\in R$
is a distributed transition over $Q$ and $S(Q')$ for some $Q, Q'\subset \Pi$.  Given such a set $R$ and $P\subset \Pi$,
$
R(P) := \{ t\in R : t \text{ is over } Q \text{ and } S(Q'), Q \cup Q'\subseteq P\}
$.
\end{definition}
Note that $Q$ and $Q'$ above are not necessarily related.  In particular, a transaction may be over agents $Q$ and states $S(Q')$ for $Q'\supset Q$.

\begin{tcolorbox}[colback=gray!5!white,colframe=black!75!black,top=4pt,bottom=4pt]
\begin{definition}[Transactions-Based Protocol]\label{definition:protocol-transactions}
Let $S$ be a local-states function and $R$ a set of transactions over $S$.
Then a \temph{protocol $\calF$ over $R$ and $S$} includes for each set of agents $P\subset \Pi$ the transactions-based distributed transition system $\calF(P)$ over $P$, $S(P)$, and $R(P)$,  $\calF(P) := (S(P)^P,\{s0\}^P,R(P){\uparrow}P)$.
\end{definition}
\end{tcolorbox}
Next we aim to find conditions under which a transactions-based protocol is grassroots.
In the original paper~\cite{shapiro2023grassrootsBA}, fulfilling three conditions were deemed sufficient: Asynchrony, interference-freedom, and interactivity.  Intuitively speaking, transactions-based distributed transition systems are asynchronous by construction, as the essence of a transaction is that it can be taken no matter what the states of non-participants are.  They are also non-interfering for the same reason.  The following Proposition captures this:

\begin{proposition}\label{proposition:oblivious}
A transactions-based protocol is oblivious. 
\end{proposition}

To prove it, we need the following Lemma:
\begin{lemma}[Closure Transitivity]\label{lemma:closure-transitivity}
Let $\emptyset \subset P \subset P' \subset \Pi$ and $R$ a set of transactions.
Then 
$$(R(P){\uparrow}P){\uparrow}P' = R(P){\uparrow}P'.$$
\end{lemma}
\begin{proof}
We argue both directions of the equality:
\begin{enumerate}
    \item $\subseteq$: Let $t\in R(P)$ and consider any $t'\in (t{\uparrow}P){\uparrow}P'$.  By definition, $t_p = t'_p$ if $p\in P$, $p$ is stationary in $t'$ if $p\in P'\setminus P$, which, by the definition of closure,  $t'\in t{\uparrow}P'$, namely $t'\in R(P){\uparrow}P'$.
    
    \item $\supseteq$:   Let $t'\in R(P){\uparrow}P'$.  Then there is a transaction  $\hat{t}\in R(P)$ over some $Q\subseteq P$, for which $t'\in \hat{t}{\uparrow} P'$. By construction, $\hat{t}_p = t'_p$ if $p\in Q$ and $p$ is stationary in $t'$ if $p\in P'\setminus Q$. 
    By definition of closure, $t=\hat{t}{\uparrow}P$ satisfies 
    $t'\in t{\uparrow}P'$, thus $t'\in (\hat{t}{\uparrow}P){\uparrow}P'$, namely  $t'\in (R(P){\uparrow}P){\uparrow}P'$.
\end{enumerate}
\end{proof}

We can now prove the Proposition:
\begin{proof}[Proof of Proposition \ref{proposition:oblivious}]
Let $\calF$ be a protocol over the state function $S$, $R$ a set of transactions over $S$, and $\emptyset \subset P \subset P' \subseteq \Pi$.  We have to show that $T(P){\uparrow}P'\subseteq T(P')$.
By definition $T(P)=R(P){\uparrow}P$ and $T(P')=R(P'){\uparrow}P'$,  thus:  
$$
T(P){\uparrow}P' = (R(P){\uparrow}P){\uparrow}P' = R(P){\uparrow}P'  \subseteq R(P'){\uparrow}P' =T(P').
$$
by definition of $T(P)$, Lemma \ref{lemma:closure-transitivity}, and since $P\subset P'$ and therefore $R(P)\subseteq R(P')$.
\end{proof}

The remaining condition is being interactive, which we aim to capture as follows:

\begin{definition}[Interactive Transactions]\label{definition:interactive}
A set of transactions $R$ over a local-states function $S$ is \temph{interactive} if for every $\emptyset \subset P \subset P' \subset \Pi$ and every configuration $c\in C(P')$ such that $c/P\in C(P)$, 
there is a computation $(c\xrightarrow{*} c')\subseteq R(P'){\uparrow}P'$ for which $c'/P\notin C(P)$.
\end{definition}

In other words, with an interactive set of transactions, any group of agents that have been so far self-contained will always have a computation with non-members that will take the group outside of its ``comfort zone'', resulting in members of the group having a local state with ``alien traces'' that could have been produced only by interacting with non-members.  

\begin{proposition}\label{proposition:interactive}
A protocol over an interactive set of transactions is interactive. 
\end{proposition}
\begin{proof}
Let $S$ be a local-states function, $R$ an interactive set of transactions over $S$, and $\calF$ 
a transactions-based protocol over $R$ and $S$,  $\emptyset \subset P \subset P' \subseteq \Pi$, and $c\in C(P')$ a configuration such that  $c{/}P\in C(P)$.
By $R$ being interactive (Definition \ref{definition:interactive}), there is a computation  $c\xrightarrow{*} c'\subseteq R(P'){\uparrow}P'$ for which $c'{/}P\notin C(P)$.  By the definition of $\calF$ as a transactions-based protocol over $R$ and $S$,  $T(P')= R(P'){\uparrow}P'$, hence this computation is of $T(P')$, establishing that $\calF$ is interactive.
\end{proof}

Our main result follows from the definitions and results above:
\begin{tcolorbox}[colback=gray!5!white,colframe=black!75!black,top=4pt,bottom=4pt]
\begin{theorem}\label{theorem:interactive-grassroots}
A protocol over an interactive set of transactions is grassroots.
\end{theorem}
\end{tcolorbox}
\begin{proof}
Let $\calF$ be a protocol over a set of transactions $R$ (Definition \ref{definition:protocol-transactions}), where $R$ is interactive (Definition \ref{definition:interactive}).
Since $\calF$ is a transactions-based protocol then, according to Proposition \ref{proposition:oblivious}, $\calF$ is oblivious.  And since $\calF$ is over an interactive set of transactions then, according to Proposition \ref{proposition:interactive}, $\calF$ is interactive.   
Therefore, by Definition \ref{definition:grassroots},  $\calF$ is grassroots..
\end{proof}

\subsection{The Three Platforms are Grassroots as Specified}

We argue briefly that the transactions-based specification of our three grassroots platforms of interest---social networks, cryptocurrencies, and democratic federations---are all interactive, and therefore according to Theorem \ref{theorem:interactive-grassroots} the protocols that they specify are all grassroots.

\begin{corollary}\label{corollary:GSN}
Grassroots Social Networks are grassroots.
\end{corollary}
\begin{proof}
Consider $P\subset P'$ and a configuration $c\in C(P')$ such that $c/P\in C(P)$.  This means that all friendships of members of $P$ in $c$ are with other members in $P$.  Hence the transaction in $R(P')$ in which a member of $P$ establishes friendship with
a member of $P'\setminus P$ satisfies the definition of interactivity.
\end{proof}

\begin{corollary}\label{corollary:GC}
Grassroots Cryptocurrencies are grassroots.
\end{corollary}
\begin{proof}
Consider $P\subset P'$ and a configuration $c\in C(P')$ such that $c/P\in C(P)$.  This means that all coins held by members of $P$ in $c$ are $P$-coins.  Hence the transaction in $R(P')$ in which a member of $P$ swaps coins with
a member of $P'\setminus P$ (possibly preceded by transactions in which the two members mint coins, if they have not done so already) satisfies the definition of interactivity.
\end{proof}

\begin{corollary}\label{corollary:GF}
Grassroots Federations are grassroots.
\end{corollary}
\begin{proof}
Consider $P\subset P'$ and a configuration $c\in C(P')$ such that $c/P\in C(P)$.  This means that the subgraph of $G$ projected by members of $P$ is a connected component of $G$. Hence the Join transaction in $R(P')$ in which a member of $P$ forms an edge with a member of $P'\setminus P$, the direction of which depends on the order $\succ$ of their identifiers, satisfies the definition of interactivity.
\end{proof}
\section{Implementation}\label{section:implementation}

The notion of implementations among distributed transition systems, as well as their fault-tolerance, has been studied extensively~\cite{abadi1993composing,lamport1999specifying,menezes1995refinement,manolios2003compositional,shapiro2021multiagent,wilcox2015verdi}.
Here, we discuss this notion briefly and informally, and plan follow-on work to do so formally.

\mypara{Binary transactions} A standard way to realize binary transactions using unary transition systems is for one agent, say $p$, to \textsc{offer} the transaction to $q$, who may respond with \textsc{accept},
upon which $p$ may respond with \textsc{commit}, upon which the offered transaction is deemed to have been executed, or \textsc{abort}.
Agent $p$ may also issue \textsc{abort} before or after receiving any response from $q$ to its offer, provided $p$ has not previously issued \textsc{commit}. 

A challenge in this implementation is that a faulty $p$ may fail to either \textsc{commit} or \textsc{abort} following an \textsc{accept} by $q$, leaving $q$ in limbo, at least in regards to this transaction.  Solutions to this are a subject of future work.

For now, we note that, worst case, a friendship offer by $p$ accepted by $q$, or an offer by  community $p$ to join community $q$, would remain in limbo.  If it is committed by $p$ at some later point, which is not convenient to $q$, then $q$ can promptly unfriend $q$ or remove $q$, as the case may be, with little or not harm done.
In case of grassroots cryptocurrencies, a swap transaction in limbo may tie coins offered by $q$, which may or may not be harmful to $q$  (not harmful if these are $q$-coins, which $q$ may mint as it pleases; or $p$-coins that $q$  tries to redeem, and if $p$ is non-responsive it might indicate that $p$-coins are not worth much anyhow).

\mypara{$k$-ary transactions}
The $k$-ary transaction, in which two communities joined or one community leaves another, calls for a consensus protocol executed by the members of the community.
To a first approximation, any permissioned consensus (Byzantine Atomic Broadcast) protocol would do.  However, the transactions entail changes in the set of agents that participate in consensual decisions, known as \emph{reconfiguration} in the consensus literature~\cite{dolev1997dynamic,aguilera2011dynamic,spiegelman2016dynamic}.
More generally, the grassroots platform entails multiple dynamically-changing, partially-overlapping, sets of agents engaged in running partially-dependent consensus protocols.  This challenge seems to have not been previously addressed in a principled way.

\section{Related and Future Work}\label{section:conclusion}

Atomic transactions have been investigated early in distributed computing, mostly in the context of database systems~\cite{lampson1981chapter,lynch1993atomic,lynch1988theory}.  Most research since and until today focuses on their efficient and robust implementation~\cite{bravo2019reconfigurable,chockler2021multi}.  The integration of atomic transactions in programming languages has also been explored~\cite{borgstrom2009compositional}).
In terms of formal models of concurrency, the extension of CCS with atomic transactions has been investigated in the past~\cite{acciai2007concurrent,de2010communicating,de2010liveness}, but without follow-on research, so it seems.  While transition systems have been the bedrock of abstract models of computation since the Turing machine, 
we are not aware of previous attempts to explore atomic transactions within their context.

Previous work on formal implementations of grassroots platforms employed unary transition systems~\cite{shapiro2021multiagent,shapiro2023gsn,shapiro2024gc,lewis2023grassroots}.  
While this new definition of a grassroots protocol tries to capture the same intuition as the original one~\cite{shapiro2023grassrootsBA}, the new definition is crisper.  It is also more restrictive in two senses, and more lax in a third:
\begin{enumerate}
    \item It is specified in terms of configurations and transitions not runs, so its restriction applies to all configurations, not only those reachable via a run.
    \item A technical limitation of comparing sets of runs, as done in the original definition, is that doing so does not capture the internal/hidden nondeterminism of intermediate configurations.  So, according to the original definition, the smaller group $P$ may have a run that leads to a configuration with multiple choices, while the larger group $P'$ has a set of runs, each leading separately to only one of these choices of $P$, but no run of $P'$ leads to a configuration in which the members of $P$ have the same choices they would have if they were to run on their own.   The current definition in terms of configurations and transitions, rather than sets of runs, eliminates this deficiency. 
    \item On the other hand, the original definition in terms of runs also addressed liveness.  Within the original definition, an all-to-all dissemination protocol $\calF$ in which $\calF(P)$ satisfies the liveness requirement that every message sent by an agent in $P$ eventually reaches every agent in $P$, is not grassroots.  The reason is that a run of $P$, which is live when member of $P$ run on their own, being oblivious of members outside $P$, would not be live within the context of $P'$, as it would  not provide dissemination between members of $P$ and members of $P'\setminus P$, indefinitely.  We consider this limitation of the new definition, as it relates to all-to-all dissemination, hypothetical, as it seems that such a protocol could not be realized without global directory (e.g., a DHT) for agents to find each other, which would not be oblivious, and hence not grassroots also under the new definition. 
\end{enumerate}
While liveness is well-understood in the context of unary transition systems, it is more opaque in the more abstract case of atomic transactions, as it may not be possible to identify the culprit in case an enabled transaction is never taken.  One the other hand, a unary implementation of atomic transactions clearly has liveness requirements. Hence, we opted not to incorporate liveness in the current context of atomic transactions, and will revisit it when the unary implementation of these specification---a subject of future research---is addressed.



\bibliography{bib}



\end{document}